\documentclass[journal]{IEEEtran}
\usepackage[pdftex]{graphicx} 
\usepackage{tikz}
\usetikzlibrary{positioning}
\usetikzlibrary{shapes,arrows}
\usetikzlibrary{shadows.blur}
\usepackage{tikz,times}
\usetikzlibrary{mindmap,backgrounds}
\usepackage[]{algorithm2e}
\usepackage[]{algorithmic}
\usepackage{mathrsfs}

\usepackage{epstopdf} 
\usepackage{cite}
\usepackage{subfig}
\usepackage{eurosym}
\usepackage{multirow}
\usepackage{amssymb}
\usepackage{amsmath}
\definecolor{armygreen}{rgb}{0.35, 0.50, 0.30}
\usepackage{amsfonts}
\usepackage{arydshln}
\usepackage{color}
\usepackage{smartdiagram}
\usetikzlibrary{matrix,arrows.meta}
\definecolor{orange}{rgb}{1,0.5,0}

\usepackage{layout}
\usepackage{hhline}
\usepackage{siunitx}

\usepackage{nomencl}
\makenomenclature

\usepackage{cases}
\usepackage{dblfloatfix}
\usepackage{tabularx,booktabs}
\newcolumntype{C}{>{\centering\arraybackslash}X} 
\usepackage{lipsum}
\usepackage{float}
\usepackage{hyperref}
\usepackage{tabu}
\usepackage{boldline,multirow}
\usepackage[T1]{fontenc}
\usepackage[11pt]{moresize}
\usepackage{array}
\newcolumntype{P}[1]{>{\centering\arraybackslash}p{#1}}
\makeatother
\usepackage{verbatim}
\usetikzlibrary{shapes,arrows, chains}
\usetikzlibrary{positioning, calc}
\usepackage{caption}
\usepackage{bm}
\usepackage{breqn}
\usepackage{tikz,pgfplots}
\usepackage{epsfig}
\usepackage{amsbsy}

\usepackage{breqn}
\usetikzlibrary{mindmap,backgrounds}

\usepackage{mathtools}

\DeclarePairedDelimiter\floor{\lfloor}{\rfloor}

\tikzfading[name=priorityarrowfadingdown,
top color=transparent!5,
bottom color=transparent!80
]

\tikzset{priority
arrow fill/.style={
  fill=gray,
  path fading=priorityarrowfadingdown
  }
}

\makeatletter
\NewDocumentCommand{\smartdiagramx}{r[] m}{
    \StrCut{#1}{:}\diagramtype\option
    \IfStrEq{\diagramtype}{priority descriptive diagram}{
        \pgfmathparse{subtract(\sm@core@priorityarrowwidth,\sm@core@priorityarrowheadextend)}
        \pgfmathsetmacro\sm@core@priorityticksize{\pgfmathresult/2}
        \pgfmathsetmacro\arrowtickxshift{(\sm@core@priorityarrowwidth-\sm@core@priorityticksize)/2}
        \begin{tikzpicture}[every node/.style={align=center,let hypenation}]
        \foreach \smitem [count=\xi] in {#2}{\global\let\maxsmitem\xi}
        \foreach \smitem [count=\xi] in {#2}{%
            \edef\col{\@nameuse{color@\xi}}
            \node[description,drop shadow](module\xi)
            at (0,0+\xi*0.65*\sm@core@descriptiveitemsysep) {\smitem};
            \draw[line width=\sm@core@prioritytick,\col]
            ([xshift=-\arrowtickxshift pt]module\xi.base west)--
            ($([xshift=-\arrowtickxshift pt]module\xi.base west)-(\sm@core@priorityticksize pt,0)$);
        }%
        \coordinate (A) at (module1);
        \coordinate (B) at (module\maxsmitem);
        \CalcHeight(A,B){heightmodules}
        \pgfmathadd{\heightmodules}{\sm@core@priorityarrowheightadvance}
        \pgfmathsetmacro{\distancemodules}{\pgfmathresult}
        \pgfmathsetmacro\arrowxshift{\sm@core@priorityarrowwidth/2}
        \begin{pgfonlayer}{background}
        \node[priority arrow,rotate=180,transform shape] at ([xshift=-\arrowxshift pt]module\maxsmitem.north west){};
        \end{pgfonlayer}
        \end{tikzpicture}
    }{}
}%

\newcommand{\cone}{[C1]}
\newcommand{\ctwo}{[C2]}
\newcommand{\cthree}{[C3]}
\newcommand{\cfour}{[C4]}

\makeatother

\begin{document}
\title{Solver-Free Heuristics to Retrieve Feasible Points for Offshore Wind Farm Collection System}
\author{Juan-Andr\'es~P\'erez-R\'ua
\thanks{Juan-Andr\'es~P\'erez-R\'ua is with DTU Wind Energy, Frederiksborgvej 399, 4000 Roskilde, Denmark (e-mail: juru@dtu.dk)}
\thanks{}}
\markboth{}%
{Shell \MakeLowercase{\textit{et al.}}: Bare Demo of IEEEtran.cls for IEEE Journals}
\maketitle
\begin{abstract}
A set of solver-free heuristics for the offshore wind collection system problem are presented. Currently, methods of this type are not able to cope with typical constraints, and most of their variations minimize only for accumulated cable length. The first algorithm is a two-steps decision process, where the output is the design of a tree network satisfying cable thermal limits constraints, but vulnerable to violate planarity constraints. Subsequently, a cable crossings repair heuristic is introduced in order to fix infeasible points from the first heuristic. Finally, a refining heuristic (negative cycle cancelling refining heuristic) takes over to improve feasible points. The latter iteratively swaps cables, intending to find cycles with negative costs that will lead to investment savings. The sequence of heuristics supports the most important restrictions of the problem. The applicability of the workflow is empirically demonstrated by means of a set of large-scale real-world offshore wind farms. The numerical results indicate that: (i) feasible points can be retrieved in computing times in order of seconds, and (ii) warm-starting can help solvers to converge significantly faster for problems with solution time in order of several hours.
\end{abstract}
\begin{IEEEkeywords}
Offshore wind , Minimum cost flow, Heuristics, Global optimization, Integer programming, Medium voltage collection system network.
\end{IEEEkeywords}
\IEEEpeerreviewmaketitle
\printnomenclature

\section{Introduction} 
\label{introduction}
\IEEEPARstart{C}{ost} reductions for renewable energy generation is on the top of political agendas, with the objective of support the worldwide proliferation of these systems. Subsidy-free calls become more frequent, as is the case for offshore wind auctions in Germany since 2017 and in Netherlands since 2018, or in China for onshore wind from 2021 \cite{GWEC2020Global2019}. Regarding offshore wind, its steep evolution during the last 12 years, where from 2009 to 2019 moved from being $1\%$ to $10\%$ of the global wind installations \cite{GWEC2020Global2019b}, is a strong proof of the maturity of the industry. The potential of offshore wind to contribute for a successful green energy transition within the right time is clear and understood by authorities and industry.

The Balance of Plant (BoP) to support the installation and operation of Wind Turbines (WTs) for Offshore Wind Farms (OWFs) can be broken down as: submarine cables, offshore substations (OSSs), converter stations for direct current technology, foundations, structures, and control equipment. BoP represents around $30\%$ of the overall levelized cost of energy (LCoE), with electrical systems being around $15\%$ \cite{ORECatapult2020WindCosts}.

The OWF collection system problem is defined as the design of the medium voltage network to interconnect WTs towards OSSs. It has been studied with particularly increased attention over the past ten years \cite{Perez-Rua2019ElectricalReview,Lumbreras2013OffshoreReview}. Finding the global optimum of this problem is generally NP-hard \cite{Jothi2005ApproximationDesign}. Three fundamental clusters of methods for tackling this problem are: heuristics, metaheuristics, and global optimization.

Global optimization encompasses several modelling options, like Binary Integer Programming (BIP) \cite{Cerveira2016OptimalCase},  Mixed Integer Linear Programming (MILP) \cite{perez2019global,Bauer2015TheProblem, Fischetti2018OptimizingLosses,Fischetti2018MixedRouting,Berzan2011AlgorithmsFarms,Wedzik2016AOptimization}, MILP with decomposition techniques for stochastic programming \cite{Lumbreras2013AProblems,Lumbreras2013OptimalStrategies}, Mixed Integer Quadratic Programming (MIQP) \cite{Hertz2017DesignAvailable,Banzo2011StochasticFarms}, and Mixed Integer Non-Linear Programming (MINLP) \cite{Chen2016CollectorFarms,Pillai2015OffshoreOptimization}. In general, external solvers based on branch-and-cut method are used to solve these formulations. Therefore, in this context, solver-free methods are defined by approaches with alternative mechanisms or set of policies, like  stochastic metaheuristics, such as genetic algorithm \cite{Gonzalez-Longatt2012OptimalApproach} or swarm optimization \cite{Hou2016OptimisationMethod}, and deterministic heuristics, as Prim \cite{Hou2016OptimisationMethod}, open vehicle routing \cite{Bauer2015TheProblem}, or Esau-Williams \cite{sanchez2018multidisciplinary,wade2019investigation,Amaral2017OffshoreLosses}. The biggest advantage of heuristics over metaheuristics is their faster convergence, which make them suitable for finding prompt solutions or for co-optimization \cite{sanchez2018multidisciplinary}.

This manuscript focuses on solver-free heuristics due to their fast convergence and capability to combine with other methods, as during pre-feasibility stage, many OWF collection system designs must be carried out, accounting for different parameters with associated uncertainty. Likewise, provided a feasible point, modern branch-and-cut solvers could improve their operation implementing a warm-starting strategy. 

While fast, the heuristics proposed in the literature to address this problem present these disadvantages: (i) they struggle to satisfy typical engineering constraints of the OWF collection system problem (non-redundant topology and no cable crossings), and (ii) they minimize for total length, and not directly the initial investment \cite{Perez-Rua2019ElectricalReview}. For the second aspect, to the best of the author's knowledge, only the sequence of works \cite{gritzbach2018towards,gritzbach2019engineering,gritzbach2020negative} have proposed heuristics to iteratively minimize total investment of the wind farm collection system. The algorithms are inspired by minimum cost flow theory, where different strategies are proposed to increase likelihood of obtaining global minimum within very short computing times. Nevertheless, those algorithms do not explicitly support basic engineering constraints for the OWF collection system.

In this sense, the main contributions of this manuscript are: (i) propose solver-free heuristics that enforce the satisfiability of relevant constraints for the problem's nature, and (ii) improve the performance of external solvers when using global optimization models with optimality certificate. 

In Section \ref{heuristics}, the heuristics are deployed with theoretical elaboration. Section \ref{global} formulates the global optimization model. Computational experiments are performed in Section \ref{experiments}, and conclusions are stated in Section \ref{conclusion}.

\section{Problem Definition and Modelling}
\label{prom_model}
The aim is to design the collection system electrical network for an OWF, i.e., to interconnect the $n_{\text T}$ WTs to the available OSSs, $n_{\text S}$, using a list $\mathcal{C}$ of cables available, while minimizing the total investment cost \cite{perez2019global,Perez-Rua2020sus}. The electrical network is represented as a static problem with respect to time with no redundancies (i.e. a forest), and the nominal power being generated by the WTs.

\sloppy Let the OSSs and WTs define the sets $\mathcal{S} = \left\lbrace1,\cdots,n_{\text S}\right\rbrace$ and $\mathcal{T} = \left\lbrace 1+n_{\text S},\cdots,n_{\text S}+n_{\text T}\right\rbrace$, respectively. The full set of points is denoted as $\mathcal{N} =\mathcal{S}\cup\mathcal{T}$, i.e., each point is assigned a unique natural number identifier. The Euclidean distance between points $i\in\mathcal{N}$ and $j\in\mathcal{N}$, is denoted as $d_{ij}$. The weighted directed graph $\mathcal{G}=(\mathcal{N},\mathcal{A},\mathcal{D})$ gathers all relevant graph-related parameters, where $\mathcal{N}$ represents the vertex set, $\mathcal{A}$ the set of arcs arranged as a pair-set $a\in\mathcal{A}:a=(i,j)$, and $\mathcal{D}$ the set of distances $d_{a}$. 

In order to simplify the problem, $\mathcal{A}$ may be truncated to only contain the arcs linked to the nearest $\upsilon<n_{\text T}$ WTs to each WT (shortest distance), plus the arcs from WTs to OSSs. Similarly, inverse arcs from OSSs to WTs should be neglected due to the flow direction. Likewise, as flow between OSSs is, in practical terms, forbidden, all arcs stemming from $\mathcal{S} \times \mathcal{S}$ are eliminated as well.

Let $n_{\text C}$ cables be available. The capacity of a cable $c \in \mathcal{C}$ is $q_c$ measured in terms of maximum number of WTs supported downstream (with respect to flow towards the OSSs). Furthermore, let $\mathcal{Q}$ be the set of thermal capacities sorted as in $\mathcal{C}$ (non-decreasing order). Each cable type $c$ has a cost per unit of length, $w_{c}\in\mathcal{W}$, in such a way that $\mathcal{C}$, $\mathcal{Q}$, and $\mathcal{W}$ are all comonotonic.

Generally, a standard feasible collection system design includes the following engineering constraints \cite{perezwes}: 
\begin{description}
    \item[\cone] (Hard) A tree topology must be enforced. This means that there must be only one electrical path from each WT towards a OSS.
    \item[\ctwo] (Hard) The thermal capacity of cables must not be exceeded.
    \item[\cthree] (Hard) Cables must not lay over each other (no crossing cables) due to practical installation aspects.    
    \item[\cfour] (Soft) The number of main feeders, i.e., cables reaching directly the OSS, might be limited to a maximum $\phi$.
\end{description}
Constraints \cone\ and \ctwo\ define a canonical computer science problem, known as Capacitated Minimum Spanning Tree (C-MST) \cite{canonical}, a NP-Hard problem. On the top of the previous two constraints, planarity constraint \cthree\ is forced due to practical limitation aspects present during the construction stage of this type of projects. Finally, spatial constraint \cfour\ is generally not binding for a large enough maximum capacity  $Q=\max \mathcal{Q}$, therefore in most of the cases is relaxed as done in this manuscript. \cfour\ is deemed as a soft constraint,\footnote{In operations research language, hard constraints delimit the feasible set, while soft constraints define the objective function, which in this case is equal to total investment.} and its implication is associated to the adaptable physical properties of main switchgears.
\section{Solver-Free Heuristics}
\label{heuristics}
\subsection{Two-steps Heuristic (TSH): C-MST and Cable Assigning}
\label{two_steps}
\begin{figure}[H]
  \centering \includegraphics[width=0.5\textwidth]{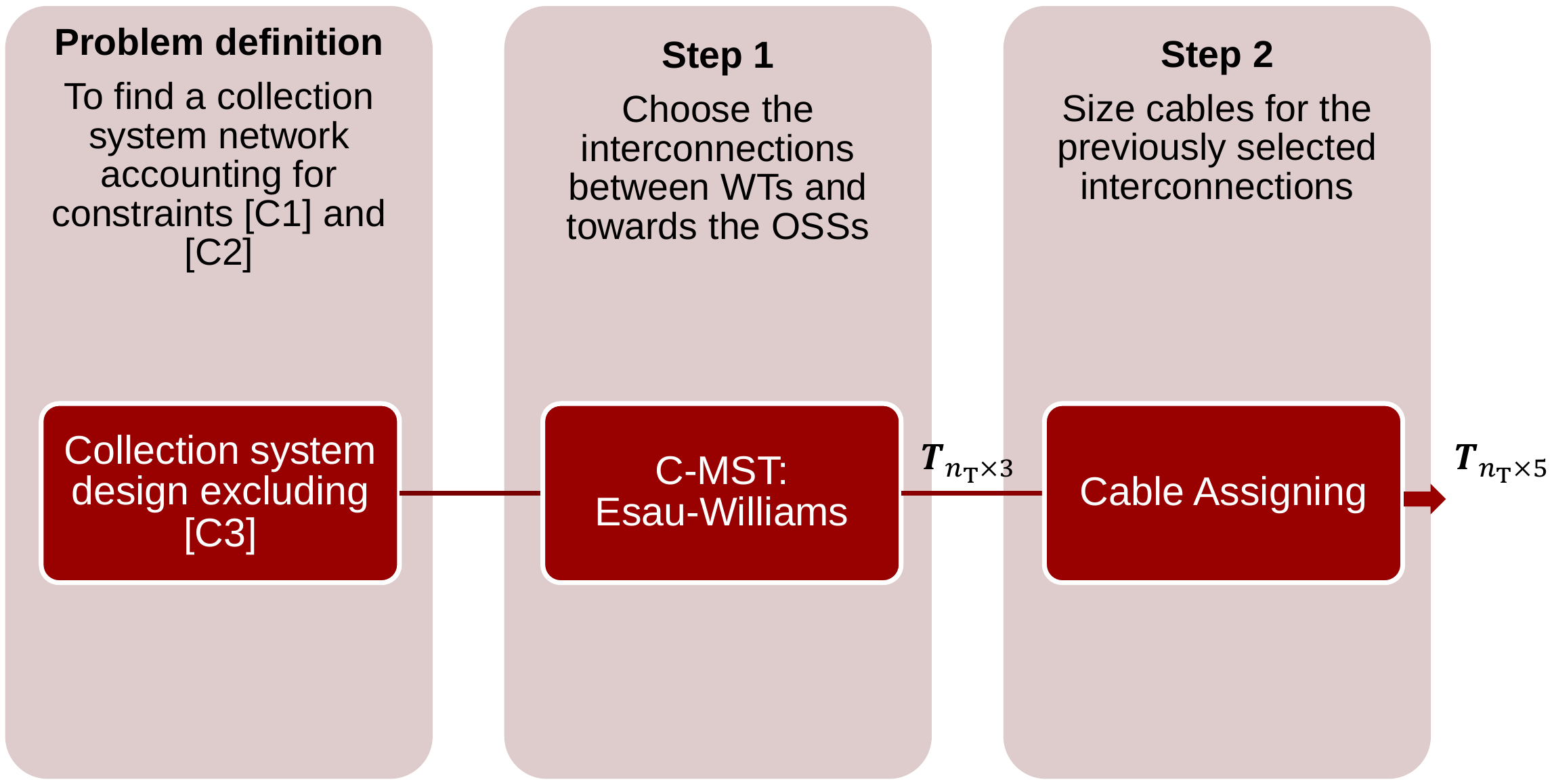}
  \caption{Flowchart of the two-steps heuristic for the offshore wind collection system problem}
  \label{fig:two_steps}
\end{figure}

A first approach to tackle the problem described in Section \ref{prom_model} is a two-steps heuristic \cite{perezwes} (see Fig. \ref{fig:two_steps}). The first step of this algorithm consists in determining the connections topology of the network, i.e., to activate edges,\footnote{Note that in contrast to the problem definition of Section \ref{prom_model}, in this case an edge $[i,j]$ has no directionality and represents both arcs $(i,j)$ and $(j,i)$.} between the nodes in $\mathcal{N}$ without sizing cables. Classic C-MST algorithms such as Prim \cite{Prim1957ShortestGeneralizations}, Kruskal \cite{Kruskal1956OnProblem}, and  Esau-Williams \cite{Esau1966OnNetwork} support constraints \cone\ and \ctwo, however they usually fail when introducing \cthree. These algorithms lead to equal solutions when constraints are not binding, as they inherently follow the same underlying mechanism based on trade-off values calculated from edges lengths and nodes weights \cite{Kershenbaum1974ComputingEfficiently}. Numerous computational experiments demonstrate the better performance of Esau-Williams heuristic, in terms of solution quality (for binding constraints) and likelihood to satisfy \cthree, compared to the other algorithms \cite{perez2019heuristics}. The output of Step 1 is a matrix $\boldsymbol{T}$ where each row represents an edge, with columns defining the connected nodes and connection length.

The cable assigning algorithm (Step 2) is based on the calculation of the number of downstream WTs connected through every edge of the network obtained in Step 1. This is followed up with the subsequent selection of the cheapest cable type for each active edge. The first task is achieved employing an embedded transversing algorithm (depth first search), which explores the undirected graph stemming from $\boldsymbol{T}$. The starting points are the nodes in $\mathcal{S}$ (OSSs). By means of this exploration, the order of nodes is re-arranged with directionality from roots (OSSs) towards leaves (WTs). In addition to the columns of $\boldsymbol{T}$ from Step 1, information regarding the the number of downstream WTs for each edge, and cheapest cable type supporting those generator units, are concatenated and presented in the final $\boldsymbol{T}$.    
\subsection{Cable Crossings Repair Heuristic (CCRH)}
\label{fixing}
The edges matrix $\boldsymbol{T}$ after Step 2 in Fig. \ref{fig:two_steps} most likely does not satisfy \cthree. If that is not the case, then luckily a feasible point for the collection system electrical network has been found out. Otherwise, Algorithm \ref{alg:alg1} is required in order to (hopefully) eliminate all cable crossings. The algorithm's mechanism is to swap infeasible edges by others which do not introduce violations to the design in a sequential deterministic manner. 

The process starts in line 6, where a list of crossing edges $\boldsymbol{Crossings}$ is generated, each row containing all edges crossing a given edge. After sorting the list for prioritizing the elimination of the largest number of crossings at once, in line 7, the algorithm is succesfully terminated in line 9 as no cables are crossing with each other, returning a crossing-free matrix $\boldsymbol{T}$. Contrarily, if after exhausting all sequence of trials there is at least one crossing, the process is stopped in line 13, with outcome stamped as infeasible (guaranteeing termination).

In line 15 a row of $\boldsymbol{Crossings}$ is selected, for which a potential edge to eliminate $Eliminate$ is fetched (line 16). A matrix $\boldsymbol{T}_{\text p}$ equal to $\boldsymbol{T}$ but with $Eliminate$ deleted is obtained in line 17. In the next two lines, those nodes out of the network due to the edge deletion are stored in $\boldsymbol{Nodes}$, and a candidate edges matrix $\boldsymbol{Candidates}$ to integrate them back into the system are created, respectively. $\boldsymbol{Candidates}$ can not contain any potential edge overlapping with the existing topology in $\boldsymbol{T}_{\text p}$.
\begin{algorithm}[ht]
\small
\caption{Pseudocode of the cable crossings repair heuristic for the offshore wind collection system problem}
\label{alg:alg1}
\algsetup{linenosize=\normalsize, linenodelimiter=., indent=1em}
\rule[-2ex]{0.95\linewidth}{0.5pt} 
\begin{algorithmic}[1]
\STATE{\textit{Get edge matrix, $\boldsymbol{T}$}}
\STATE{\textit{$counter1\leftarrow0$}}
\STATE{\textit{$counter2\leftarrow0$}}
\WHILE{\textit{True}}
    \IF{\textit{$counter1==0$} and \textit{$counter2==0$}}
       \STATE{\textit{Obtain list of edges crossings in $\boldsymbol{T}$, $\boldsymbol{Crossings}$}\\ (A row $l\in \boldsymbol{Crossings}$ contains the list of edges $l[1:]$ crossing with edge $l[0]$ as indexes of $\boldsymbol{T}$)}
        \STATE{\textit{Sort rows of $\boldsymbol{Crossings}$ in non-increasing order}}
        \IF{\textit{$len(\boldsymbol{Crossings})==0$}}
           \STATE{\textit{Break. All crossings eliminated}}
        \ENDIF
    \ENDIF
    \IF{\textit{$counter2==len(\boldsymbol{Crossings})$}}
       \STATE{\textit{Break. Not all crossings eliminated}}
    \ENDIF
    \STATE{\textit{$\boldsymbol{Edges}\leftarrow\boldsymbol{Crossings}[counter2]$}}
    \STATE{\textit{$Eliminate\leftarrow\boldsymbol{Edges}[counter1]$}}
    \STATE{\textit{$\boldsymbol{T}_{\text p}\leftarrow del(\boldsymbol{T},Eliminate)[:,:2]$}}
    \STATE{\textit{Find nodes out of the tree $\boldsymbol{T}_{\text p}$, $\boldsymbol{Nodes}$}}
    \STATE{\textit{Find candidate edges connecting to $\boldsymbol{Nodes}$ which do not cross with any edge in $\boldsymbol{T}_{\text p}$, $\boldsymbol{Candidates}$}}
    \STATE{\textit{$counter3\leftarrow0$}}
    \WHILE{\textit{True}}
        \STATE{\textit{$infeasible\leftarrow False$}}
        \STATE{\textit{$\boldsymbol{T}_{\text p}\leftarrow add(\boldsymbol{T}_{\text p},Eliminate,\boldsymbol{Candidates}[counter3])$}}
        \IF{\textit{Not satisfied \cone\ and \ctwo\ on $\boldsymbol{T}_{\text p}$}}
            \STATE{\textit{$infeasible\leftarrow True$}}
            \STATE{\textit{$\boldsymbol{T}_{\text p}\leftarrow del(\boldsymbol{T}_{\text p},Eliminate)$}} 
            \STATE{\textit{$counter3+=1$}}
            \IF{\textit{$counter3==len(\boldsymbol{Candidates})$}}
                \STATE{\textit{$counter1+=1$}}
                \IF{\textit{$counter1==len(\boldsymbol{Edges})$}}
                   \STATE{\textit{$counter2+=1$}}
                   \STATE{\textit{$counter1\leftarrow0$}}
                \ENDIF
                \STATE\textit{Break. Need to find new candidates}
            \ENDIF        
        \ELSE
            \STATE{\textit{Get new tree with new concatenated edge, $\boldsymbol{T}$}}
            \STATE{\textit{$counter1\leftarrow0$}}
            \STATE{\textit{$counter2\leftarrow0$}}
            \STATE\textit{Break. At least one crossing eliminated}
        \ENDIF
    \ENDWHILE
\ENDWHILE
\STATE{\textit{Return new edges matrix, $\boldsymbol{T}$}}
\STATE{\textit{Return feasibility flag, $infeasible$}}
\end{algorithmic}
\rule[1ex]{0.95\linewidth}{0.5pt}
\end{algorithm}

Lines 21 to 42 are for the inner loop in charge of exploiting matrix $\boldsymbol{Candidates}$. Line 23 incorporates to $\boldsymbol{T}_{\text p}$ the candidate edge $\boldsymbol{Candidates}[counter3]$ in the index $Eliminate$, then in line 24 is assessed if constraints \cone\ and \ctwo\ are respected in the tentatively formed network. If that is not the case, the network is infeasible, and therefore in line 26 the candidate edge $\boldsymbol{Candidates}[counter3]$ is disconsidered. The next candidate edge index by $counter3$ is retrofitted to $\boldsymbol{T}_{\text p}$, repeating the previous examination. When all alternatives in $\boldsymbol{Candidates}$ are considered, a new set of them must be computed by exploring $\boldsymbol{Crossings}$ and having a new potential to eliminate (the inner while loop is broken). The edge is only permanently eliminated if constraints \cone\ and \ctwo\ are satisfied, when this is swapped by edge $\boldsymbol{Candidates}[counter3]$ (line 37). The inner while loop is interrupted, and a new list $\boldsymbol{Crossings}$ is created, restarting the transversing counters, $counter1$ and $counter2$.

Output of Algorithm \ref{alg:alg1} is the new edges matrix $\boldsymbol{T}$ satisfying \cthree\ in case $infeasible=False$. Matrix $\boldsymbol{T}$ has same structure of obtained after the TSH. While it is impossible to formulate theoretical guarantees regarding worst-case scenario computing time and solution quality, computational experiments in Section \ref{res_rep_ref} demonstrate the effectiveness of the CCRH applied to real-world problems.

Specific computing times and likelihood of success are problem-dependent, among other factors, of the WTs and OSSs layout, list of cables available, and number of crossings after the TSH. 
\subsection{Negative Cycle Cancelling Refining Heuristic (NCCRH)}
\label{refining}
\subsubsection{Background}
\label{ncc_background}
The Negative Cycle Cancelling Algorithm (NCCA), also known as the \textit{augmenting cycle method}, was originally derived to solve a standard network problem, the \textit{Minimum Cost Flow (MCF) problem} \cite{ahuja1995network}. The classic version of the MCF problem consists in supplying the sinks from the sources by a flow $\Lambda$ in the cheapest possible way, given a directed graph $\mathcal{G}_{\text c}=(\mathcal{V}_{\text c},\mathcal{A}_{\text c},\mathcal{D}_{\text c},\mathcal{U}_{\text c},\mathcal{P}_{\text c})$, where $\mathcal{V}_{\text c}$ is the nodes set (formed by sources and sinks), $\mathcal{A}_{\text c}$ the arcs set ($a\in\mathcal{A}_{\text c}$), $\mathcal{D}_{\text c}$ the lengths set for all arcs ($d_{\text{c}_{a}}$ length of arc $a$), $\mathcal{U}_{\text c}$ the capacities set for all arcs ($u_{\text{c}_{a}}$ capacity of arc $a$), $\mathcal{P}_{\text c}$ the costs set per unit of flow for all arcs ($p_{\text{c}_{a}}$ linear cost of arc $a$). Formally, the problem can be formulated as an integer program as:
\begin{eqnarray}
\text{min} & \sum\limits_{\forall a \in \mathcal{A}_{\text c}} p_{\text{c}_{a}} \cdot \lambda_a \label{eqn:obj_mcf}\\
\text{s.t.} & f_{\Lambda}(i)=b_i \quad \forall i \in \mathcal{V}_{\text c} \label{eqn:power_bal_mcf}\\
    & 0\leq \lambda_a \leq u_{\text{c}_{a}} \quad  \forall a \in \mathcal{A}_{\text c} \label{eqn:cap_mcf}
\end{eqnarray}
The flow balance equation for node $i\in\mathcal{V}_{\text c}$ in  (\ref{eqn:power_bal_mcf}) is given as $f_{\Lambda}(i)=\sum_{\forall a_1 \in \delta^{-}(i)}\lambda_{a_1}-\sum_{\forall a_2 \in \delta^{+}(i)}\lambda_{a_2}$, where $\delta^{-}(i)$ and $\delta^{+}(i)$ are the incoming and outgoing arcs to $i$, respectively. $b_i$ is the demand at node $i$ (negative for source nodes). In  (\ref{eqn:cap_mcf}), $\lambda_a$ is an integer variable representing the flow through $a\in \mathcal{A}_{\text c}$.

The MCF problem acts as a umbrella formulation as numerous flow problems can be stated using the same concept, as for example, the transportation problem, the shortest path problem, and the maximum flow problem. 

By exploiting the complementary slackness conditions of the problem from  (\ref{eqn:obj_mcf}) to  (\ref{eqn:cap_mcf}) and its dual, the following algorithm guarantees the optimum solution (see proof of correctness in \cite{ahuja1995network}):
\begin{algorithm}[ht]
\small
\caption{Pseudocode of the classic negative cycle cancelling refining heuristic}
\label{alg:alg2}
\algsetup{linenosize=\normalsize, linenodelimiter=., indent=1em}
\rule[-2ex]{0.95\linewidth}{0.5pt} 
\begin{algorithmic}[1]
\STATE{\textit{Find a feasible point, $\Lambda$}}
\WHILE{\textit{There exists a negative cost cycle in $\mathcal{G}_{\Lambda}$}}
    \STATE{\textit{Find circuit with negative cost in $\mathcal{G}_{\Lambda}$ based on $\mathcal{P}_{\Lambda}$, $\mathcal{L}$}}
    \STATE{\textit{Find $\Delta=\min_{a\in\mathcal{L}} u_{\Lambda_{a}}$}}
    \STATE{\textit{Push $\Lambda$ on $\mathcal{L}$ with $\Delta$ units}}
\ENDWHILE
\end{algorithmic}
\rule[1ex]{0.95\linewidth}{0.5pt}
\end{algorithm}

A feasible flow $\Lambda$ satisfying (\ref{eqn:power_bal_mcf}) and (\ref{eqn:cap_mcf}) is obtained in line 1 of Algorithm \ref{alg:alg2}, for example by solving a maximum flow problem \cite{hochbaum2008pseudoflow}. The \textit{residual graph} $\mathcal{G}_{\Lambda}$, indicating how flow excess can be moved in $\mathcal{G}_{\text c}$ given the present flow $\Lambda$, is obtained in line 2.

The definition of $\mathcal{G}_{\Lambda}$ is presented in (\ref{eqn:res_graph}), where the nodes set $\mathcal{V}_{\Lambda}$ is equal to the original graph joined with the fictitious root node $i_{\text r}$, the arcs set $\mathcal{A}_{\Lambda}$ is defined as the arcs of the original graph where flow is lower than their capacity, along with the inverse arcs ($\Bar{a}$) of the original graph where flow is strictly greater than 0, plus arcs from fictitious root note $i_{\text r}$ to all nodes. $\mathcal{U}_{\Lambda}$ is called the residual capacities set, composed respectively of the remaining capacity of the arcs with flow lower than their capacity, by the flow of arcs when is greater than zero, and by infinity capacity for the arcs rooted at $i_{\text r}$. Finally, for the costs set $\mathcal{P}_{\Lambda}$, cost is equal for arcs with flow lower than capacity, negative in the inverse arcs when flow in the original graph is greater than zero, plus the costs of the arcs rooted at $i_{\text r}$ being zero.
\begin{gather} \label{eqn:res_graph}
\mathcal{G}_{\Lambda}=(\mathcal{V}_{\Lambda},\mathcal{A}_{\Lambda},\mathcal{U}_{\Lambda},\mathcal{P}_{\Lambda})\\ 
\mathcal{V}_{\Lambda}=\mathcal{V}_{\text c} \cup \left\lbrace i_{\text r} \right\rbrace \textit{ (fictitious root node)} \nonumber \\
\mathcal{A}_{\Lambda}=\left\lbrace a:a\in\mathcal{A}_{\text c} \wedge \lambda_{a}<u_{\text{c}_{a}} \right\rbrace \cup \left\lbrace a:\Bar{a}\in\mathcal{A}_{\text c} \wedge \lambda_{\Bar{a}}>0 \right\rbrace \cup \nonumber \\ \left\lbrace a:a=(i_{\text r},j) \wedge j\in\mathcal{V}_{\Lambda}\setminus \left\lbrace i_{\text r} \right\rbrace \right\rbrace \nonumber\\
\mathcal{U}_{\Lambda}=\left\lbrace u_{\text{c}_{a}}-\lambda_a:a\in\mathcal{A}_{\text c} \wedge \lambda_{a}<u_{\text{c}_{a}} \right\rbrace \cup \nonumber \\ \left\lbrace \lambda_{\Bar{a}}:\Bar{a}\in\mathcal{A}_{\text c} \wedge \lambda_{\Bar{a}}>0 \right\rbrace \cup \left\lbrace \infty: a=(i_{\text r},j) \right\rbrace \nonumber\\
\mathcal{P}_{\Lambda}=\left\lbrace p_{\text{c}_{a}}:a\in\mathcal{A}_{\text c} \wedge \lambda_{a}<u_{\text{c}_{a}} \right\rbrace \cup \left\lbrace -p_{\text{c}_{\Bar{a}}}:\Bar{a}\in\mathcal{A}_{\text c} \wedge \lambda_{\Bar{a}}>0 \right\rbrace \nonumber\\ \cup \left\lbrace 0: a=(i_{\text r},j) \right\rbrace \nonumber
\end{gather}
In line 3 of Algorithm \ref{alg:alg2} a negative cost cycle\footnote{A cycle is defined as a sequence of arcs $\left\lbrace(i,j),(j,u),(u,v),(v,i)\right\rbrace$, such that the head of an arc is equal to the tail of the next arc, and the initial node of the path is equal to the final one.} $\mathcal{L}$ (if any) must be found with initial point $i_{\text{r}}$. This is possible by means of the shortest path algorithm, Bellman-Ford \cite{bellman1958routing,ford1956network} in $\mathcal{O}(nm)$, where $n$ is the number of nodes, and $m$ the number of arcs. If a $\mathcal{L}$ is present in $\mathcal{G}_{\Lambda}$, then a surplus flow $\Delta$ (from line 4) equal to the minimum value of the residual capacities on arcs in $\mathcal{L}$ is pushed in the cycle in line 5. Ultimately, the algorithm is terminated if an $\Delta$-augmenting circuit with negative cost does not exist. During the development of the process, the incumbent is iteratively improved while always satisfying (\ref{eqn:power_bal_mcf}) and (\ref{eqn:cap_mcf}). Algorithm \ref{alg:alg2} has at most $|\mathcal{A}_{\text c}|\cdot\max{\mathcal{P}_{\text c}}\cdot\max{\mathcal{U}_{\text c}}$ iterations.
\subsubsection{Disparities of the MCF problem with the OWF collection system problem}
\label{ncc_disparities}
The NCCA provides the global optimum for the classic MCF problem. However, in spite of the similarities with the OWF collection system, the following major disparities preclude its application to this problem:
\begin{itemize}
    \item The capacities set $\mathcal{U}_{\text c}$ in the MCF problem is defined beforehand as part of the input parameters. Nevertheless, for the OWF case, this set is in fact a function of the flow, $\mathcal{U}_{\text c}(\Lambda)$. This is because of the list of cables available, $\mathcal{C}$ with capacities set $\mathcal{Q}$. 
    \item The costs set $\mathcal{P}_{\text c}$ are linear functions of the flow $\Lambda$ in the MCF problem. On the contrary in the collection system problem, the cost is a non-convex step function as illustrated in Fig. \ref{fig:cost_function}. The cost difference between consecutive steps tends to be larger than a proportional rate, but rather a polynomial or even exponential function of the power flow \cite{Lundberg2003ConfigurationParks}. In Fig. \ref{fig:cost_function}, $k$ is the number of WTs connected through an arc (equivalent to $\lambda_a$), and $g(k)$ represents the cost function for all $a\in \mathcal{A}_{\text c}$.
    \begin{figure}[H]
    \centering \includegraphics[width=0.49\textwidth]{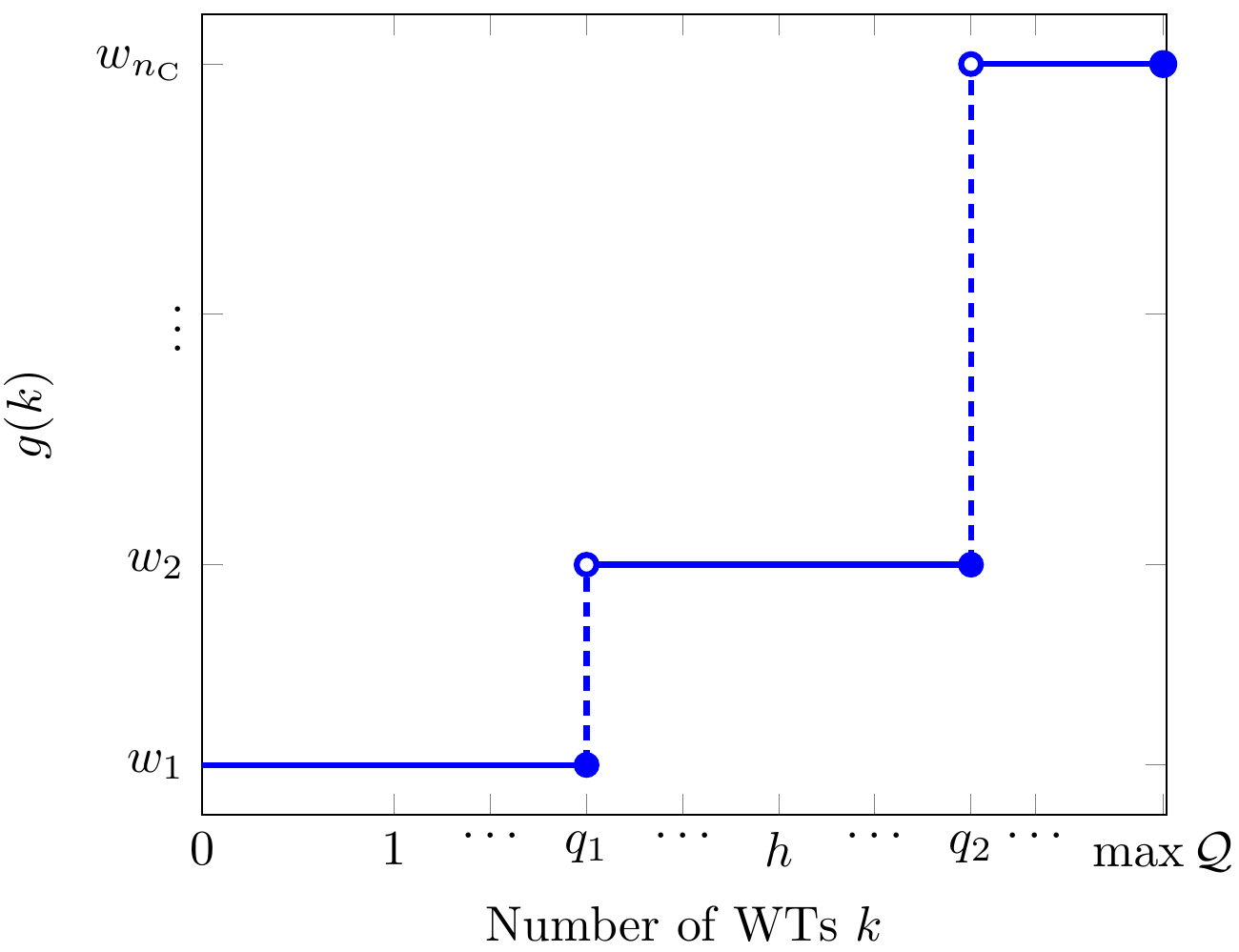}
    \caption{Connection cost function for the offshore wind collection system problem}
    \label{fig:cost_function}
    \end{figure}
    \item Constraints \cone\ (tree topology) and \cthree\ (no crossing cables) are not required in the MCF problem.  
\end{itemize}
\subsubsection{The NCCRH for the OWF collection system problem}
\label{nccrh}
Due to the main differences between the OWF collection system problem and the MCF problem stated in Section \ref{ncc_disparities}, modifications to the formulation in Section \ref{ncc_background} must be introduced. The goal is to propose a new algorithm (the NCCRH) to tackle the former problem, notwithstanding the impossibility to assure neither optimality nor theoretical successful convergence, but with a good experimental performance on real-world problems.

The input directed graph is defined as $\mathcal{G}_{\text c}=(\mathcal{V}_{\text c},\mathcal{A}_{\text c},\mathcal{D}_{\text c})$. Where $\mathcal{V}_{\text c}=\mathcal{N}$. The other two sets are in (\ref{eqn:ini_coll_owf}). The arcs set $\mathcal{A}_{\text c}$ consists of the arcs in $\mathcal{A}$ (Section \ref{prom_model}) excluding their inverses $\mathcal{A}^{-}$, and $\mathcal{D}_{\text c}$ the set of distances for each arc in $\mathcal{A}_{\text c}$.
\begin{gather} \label{eqn:ini_coll_owf}
\mathcal{A}_{\text c}=\mathcal{A}^{+}=\left\lbrace a:a\in\mathcal{A} \wedge \Bar{a}\notin\mathcal{A}^{+} \right\rbrace \\ 
\mathcal{D}_{\text c}=\left\lbrace d_a: a\in\mathcal{A}_{\text c}\right\rbrace \nonumber\\
\mathcal{A}^{-}=\left\lbrace a:\Bar{a}\in\mathcal{A}^{+} \right\rbrace \nonumber
\end{gather}
The residual graph in this case is not a function of the flow $\Lambda$, but a rather a constant network given by (\ref{eqn:res_graph_coll_owf}). Nodes set $\mathcal{V}_{\Lambda}$ includes the original set $\mathcal{V}_{\text c}$, a cluster transfer node $i_{\text o}$ to model surplus flow interchange between OSSs, and the fictitious root node $i_{\text r}$. On the other hand, arcs set $\mathcal{A}_{\Lambda}$ is composed of arcs with tail at OSS nodes $j\in\mathcal{S}$ and head at the cluster transfer node $i_{\text{o}}$, the inverse of these arcs, plus sets $\mathcal{A}^{+}$ and $\mathcal{A}^{-}$, and plus arcs connecting the fictitious root node $i_{\text{r}}$ to the rest of nodes of the residual graph.

Comparing (\ref{eqn:res_graph_coll_owf}) with (\ref{eqn:res_graph}) is noticeable that the residual capacities set $\mathcal{U}_{\Lambda}$ and costs set $\mathcal{P}_{\Lambda}$ have not been defined for this problem. As aforementioned, the network associated to $\mathcal{G}_{\Lambda}$ is constant for flow $\Lambda$. The feasibility to push a surplus flow $\Delta$ needs to be assessed with a \textit{residual cost function} $r(\lambda_a,\Delta)$, as initially proposed in \cite{gritzbach2018towards}, where the cable capacities are intrinsically accounted for. 
\begin{gather} \label{eqn:res_graph_coll_owf}
\mathcal{G}_{\Lambda}=(\mathcal{V}_{\Lambda},\mathcal{A}_{\Lambda})\\
\mathcal{V}_{\Lambda}=\mathcal{V}_{\text c} \cup \left\lbrace i_{\text o} \right\rbrace \textit{ (cluster transfer node)} \cup \left\lbrace i_{\text r} \right\rbrace \nonumber\\
\mathcal{A}_{\Lambda}=\mathcal{A}^{\text{o}^{-}} \cup \mathcal{A}^{\text{o}^{+}} \cup \mathcal{A}^{+} \cup \mathcal{A}^{-} \cup \mathcal{A}^{\text r} \nonumber\\
\mathcal{A}^{\text{o}^{-}}=\left\lbrace a:a=(j,i_{\text o}) \wedge j\in\mathcal{S}\right\rbrace \nonumber\\
\mathcal{A}^{\text{o}^{+}}=\left\lbrace a:a=(i_{\text o},j) \wedge j\in\mathcal{S}\right\rbrace \nonumber\\
\mathcal{A}^{\text r}=\left\lbrace a:a=(i_{\text r},j) \wedge j\in\mathcal{V}_{\Lambda} \setminus \left\lbrace i_{\text r} \right\rbrace\right\rbrace \nonumber
\end{gather}
In Section \ref{ncc_background}, $\Lambda$ is defined as the flow set, where each element contains the non-negative flow $\lambda_a$ for each arc $a\in\mathcal{A}_{\text c}$. Within this context, $\lambda_a$ may be equal to any real value, in such a way that, $\lambda_a\geq0$ if flow goes from node $i$ to $j$, where $a=(i,j)\in\mathcal{A}_{\text c}$, otherwise $\lambda_a<0$, if flow goes from $j$ to $i$. Let $\Lambda$ be redefined under this principle. Additionally, a mirroring flow set through each arc belonging to $\mathcal{A}_{\Lambda}$ is formalized in (\ref{eqn:flow_mirror}):
\begin{equation}\label{eqn:flow_mirror}
    \Lambda_{\text n}=\Lambda \cup \left\lbrace\lambda_a\leftarrow\lambda_b: a=\Bar{b}, b\in\mathcal{A}^{+}, a\in\mathcal{A}^{-}\right\rbrace
\end{equation}

The residual cost function $r(\lambda_a,\Delta)$ is recurrently applied to each arc $a\in\mathcal{A}_{\Lambda}$, given a potential positive surplus flow $\Delta$, and information about flow $\lambda_a$ contained $\Lambda_{\text n}$. This function is defined in (\ref{eqn:res_cost}).
\begin{equation}\label{eqn:res_cost}
r(\lambda_a,\Delta) = \begin{cases}
0, & (a\in\mathcal{A}^{\text{o}^{-}}) \vee (a\in\mathcal{A}^{\text{o}^{+}}: a= (i,j),  j\in\\ 
   & \mathcal{S}) \wedge \Delta\leq f_{\Lambda}(j)) \vee (a\in\mathcal{A}^{\text r})\\
g^{+},& a\in\mathcal{A}^{+} \wedge |\lambda_a+\Delta|\leq Q\\
g^{-},& a\in\mathcal{A}^{-} \wedge |\lambda_a-\Delta|\leq Q\\
\infty, & (a\in\mathcal{A}^{\text{o}^{+}}: a= (i,j),  j\in \mathcal{S} \wedge \Delta>\\
        & f_{\Lambda}(j))\vee (a\in\mathcal{A}^{+} \wedge |\lambda_a+\Delta|>Q) \vee\\
        & (a\in\mathcal{A}^{-}:a=(i,j), i\in\mathcal{T} \wedge |\lambda_a-\Delta|\\
        & >Q) \vee (a\in\mathcal{A}^{-}:a=(i,j), i\in\mathcal{S}\\
        & \wedge \Delta>\lambda_a) \text{.}
\end{cases}
\end{equation}
\begin{equation}\label{eqn_res1}
g^{+}=g(|\lambda_a+\Delta|)-g(|\lambda_a|)    
\end{equation}
\begin{equation}\label{eqn_res2}
g^{-}=g(|\lambda_a-\Delta|)-g(|\lambda_a|)    
\end{equation}
The residual cost function equals zero for all arcs in $\mathcal{A}^{\text{o}^{-}}$ or in $\mathcal{A}^{\text r}$, and for those arcs in $\mathcal{A}^{\text{o}^{+}}$ when the total incoming flow to the OSS is greater than or equal to $\Delta$. The latter avoids outgoing flow from the OSS. For the arcs in $\mathcal{A}^{+}$ that the absolute value of the surplus flow plus the flow through them is lower than or equal to the capacity of the biggest cable available, the residual cost is equal to $g^{+}$ (\ref{eqn_res1}). Similarly, arcs in $\mathcal{A}^{-}$ with absolute value of the flow through them minus $\Delta$ lower than or equal to the capacity of the biggest cable available, their residual cost is given by $g^{-}$ (\ref{eqn_res2}). Finally, the residual cost is set to infinite for infeasible arcs to avoid impractical situations, such as outgoing flow from a OSS, and cable thermal capacity excedance. 

Algorithm \ref{alg:alg3} presents the working principles of the NCCRH.  
\begin{algorithm}[ht]
\small
\caption{Pseudocode of the negative cycle cancelling refining heuristic for the offshore wind collection system problem}
\label{alg:alg3}
\algsetup{linenosize=\normalsize, linenodelimiter=., indent=1em}
\rule[-2ex]{0.95\linewidth}{0.5pt} 
\begin{algorithmic}[1]
\STATE{\textit{Form input direct graph using (\ref{eqn:ini_coll_owf}), $\mathcal{G}_{\text c}=(\mathcal{V}_{\text c},\mathcal{A}_{\text c},\mathcal{D}_{\text c})$}}
\STATE{\textit{Form residual graph using (\ref{eqn:res_graph_coll_owf}), $\mathcal{G}_{\Lambda}=(\mathcal{V}_{\Lambda},\mathcal{A}_{\Lambda})$}}
\STATE{\textit{Find an initial feasible point based on $\boldsymbol{T}$, $\Lambda$}}
\STATE{\textit{Get mirroring flow set using (\ref{eqn:flow_mirror}), $\Lambda_{\text n}$}}
\STATE{\textit{$\Lambda_{\text t}\leftarrow\Lambda$}}
\STATE{\textit{$counter1\leftarrow0$}}
\STATE{\textit{$\boldsymbol{\Delta}\leftarrow unique(|\Lambda|)$}}
\WHILE{\textit{True}}
    \STATE{\textit{$\Delta\leftarrow \boldsymbol{\Delta}[counter1]$}}
    \STATE{\textit{$\mathcal{P}_{\Lambda}\leftarrow\left\lbrace r(\lambda_a,\Delta):a\in\mathcal{A}_{\Lambda}\right\rbrace$}}
    \STATE{\textit{Get list of cycles with more than two arcs based on $\mathcal{P}_{\Lambda}$, $\boldsymbol{L}$}}
    \FOR{$\mathcal{L}\in\boldsymbol{L}$}
       \IF{\textit{$\mathcal{L}$ has negative cost}}
           \STATE{\textit{Push $\Lambda_{\text t}$ on $\mathcal{L}$ with $\Delta$ units}}
           \IF{\textit{Satisfied \cone\ and \cthree\ on $\Lambda_{\text t}$ using $\mathcal{G}_{\text c}$}}
              \STATE{\textit{$\Lambda\leftarrow\Lambda_{\text t}$}}
              \STATE{\textit{Update $\Lambda_{\text n}$ with (\ref{eqn:flow_mirror})}}
              \STATE{\textit{$\boldsymbol{\Delta}\leftarrow unique(|\Lambda|)$}}
              \STATE{\textit{$counter1\leftarrow-1$}}
              \STATE{\textit{Break. Flow improved}}
           \ENDIF
           \STATE{\textit{$\Lambda_{\text t}\leftarrow\Lambda$}}
       \ENDIF 
    \ENDFOR
    \STATE{\textit{$counter1+=1$}}
    \IF{\textit{$counter1==len(\boldsymbol{\Delta})$}}
        \STATE{\textit{Break. Algorithm terminated}}
    \ENDIF
\ENDWHILE
\STATE{\textit{Return feasible flow, $\Lambda$, and associated active arcs with cable selected in $\mathcal{G}_{\text c}$, $\boldsymbol{T}$}}
\end{algorithmic}
\rule[1ex]{0.95\linewidth}{0.5pt}
\end{algorithm}
Lines 1 to 2 initialize the required inputs $\mathcal{G}_{\text c}$ (input directed graph) and $\mathcal{G}_{\Lambda}$ (residual graph), respectively. This is continued by the derivation of a feasible flow $\Lambda$ by means of $\boldsymbol{T}$ (output of Algorithm \ref{alg:alg1}). In line 4 the mirroring flow set  $\Lambda_{\text n}$ is gotten, and then an auxiliary set $\Lambda_{\text t}$ is defined. The last initialization step in line 7 is to get the set of potential surplus flows $\boldsymbol{\Delta}$, which consists of the unique positive arc flows in the feasible flow set $\Lambda$. 

The job of the process between lines 8 to 29 is to swap one active arc $a$ such as $\lambda_a\neq0$ with an inactive one ($\lambda_a=0$), thus preserving the satisfiability of \cone, restricted to the no violation of \cthree. Constraint \ctwo\ is satisfied implicitly by the residual cost function (\ref{eqn:res_cost}).

Line 9 chooses one potential surplus flow $\Delta$ at the time, which is then utilized in line 10 to create the set of residual costs $\mathcal{P}_{\Lambda}$, such as $p_{\text{c}_{a}}\in\mathcal{P}_{\Lambda}$ is the residual cost of arc $a$ based on its flow $\lambda_a\in\Lambda_{\text n}$. The shortest path algorithm Bellman-Ford is implemented in line 11 in order to get a negative cycle (if any). Let the tracked negative cycle has the form $\left\lbrace(i,j),(j,u),(u,v),(v,w),(w,z),(z,v),(v,u),(u,i)\right\rbrace$. The co-existence of arc $(u,v)$ and its inverse $(v,u)$ in the cycle bring in a contradicting behaviour of the algorithm as pointed out in \cite{gritzbach2018towards}. For this reason this arc and its inverse are eliminated of the path, forming a cycles set $\boldsymbol{L}=\left\lbrace\left\lbrace(i,j),(j,u),(u,i)\right\rbrace,\left\lbrace(v,w),(w,z),(z,v)\right\rbrace\right\rbrace$.

A cycle $\mathcal{L}\in\boldsymbol{L}$ is selected, and in case it has a total negative cost (otherwise try another cycle within the same potential surplus flow or with the next in $\boldsymbol{\Delta}$), then $\Delta$ units of surplus flow are pushed on $\mathcal{L}$ in the flow set $\Lambda_{\text t}$ (line 14). Pushing surplus flow means in this context that if $a\in\mathcal{L}\wedge a\in\mathcal{A}^{+}$, then the flow is pushed \textit{forward} through $a$, i.e., $\lambda_a\leftarrow\lambda_a+\Delta$, or contrarily, if $a\in\mathcal{L}\wedge\Bar{a}\in\mathcal{A}^{+}$, then the flow is pushed \textit{backwards} through $\Bar{a}$, i.e., $\lambda_{\Bar{a}}\leftarrow\lambda_{\Bar{a}}-\Delta$. This guarantees the flow conservation.

Constraints \cone\ and \cthree\ are evaluated for the network stemming from arcs with flow different than zero in $\Lambda_{\text t}$ according to $\mathcal{G}_{\text c}$. If both constraints are satisfied, then $\Lambda$, $\Lambda_{\text n}$, $\boldsymbol{\Delta}$, and $counter1$ are reset along with the whole previous process. Otherwise, the next cycle $\mathcal{L}$ is studied. The algorithm is terminated in line 27, when all the potential surplus flows have been exhausted. This secures termination of the algorithm. The output is a (possibly) improved flow $\Lambda$ and associated improved edges matrix $\boldsymbol{T}$.

The aim is to run the NCCRH in a time in order of seconds, with worst-case scenario of not improving the feasible point after the CCRH, due to the hardness of simultaneously considering \cone\ and \cthree. 

\section{Global Optimization Model: A MILP Program}
\label{global}
The following formulation is based on that proposed in \cite{perez2019global} with some adaptations to fit the problem defined in Section \ref{prom_model}. Solving a MILP program by a state-of-the-art branch-and-cut solver brings along the advantage of providing a solution with optimality certificate, known as the GAP \cite{Wolsey2014IntegerOptimization}. The GAP is defined as the percentage by which the best known achievable solution value is distant to the best known feasible point. A GAP of $0\%$ means $100\%$ of confidence in the finding of the global minimum.
\subsection{Variables}
\label{var}
Let $x_{ij}$ represent a binary variable that is one if the arc between the vertex $i$ and $j$ is selected in the solution, and zero otherwise. Likewise, the binary variable $y^{k}_{ij}$ models the $k$ number of WTs connected downstream from $i$, including the WT at node $i$ (under the condition that $x_{ij} = 1$). The possible maximum value of $k$ for $j\in\mathcal{S}$ is equal to $h(j)=Q$, while for $j\in\mathcal{T}$ is $h(j) = Q-1$. This means that the biggest cable available could be only used at maximum capacity when is connected from a OSS. 
\subsection{Objective Function}
\label{obj}
The linear objective function of the mathematical model is 
\begin{equation} \label{eqn:objective}
\min{\sum\limits_{i \in \mathcal{T}}\sum\limits_{j \in \mathcal{N}:j\neq i}\sum\limits_{k = 1}^{h(j)}  g(k) \cdot y^{k}_{ij}}
\end{equation}

By means of the definition of $g(k)$ as illustrated in Fig. \ref{fig:cost_function}, the constraint \ctwo\ is implicitly satisfied. 
\subsection{Constraints}
\label{cons}
To simultaneously ensure a tree topology and to define the head-tail convention, the next expression is included into the MILP model
\begin{equation} \label{eqn:c1}
\sum\limits_{j \in \mathcal{N}:j\neq i}\sum\limits_{k = 1}^{h(j)} y^{k}_{ij} = 1 \quad  \forall i\in\mathcal{T}
\end{equation}

The flow conservation, which also avoids disconnected solutions, is considered by means of one linear equality per WT:
\begin{equation} \label{eqn:c2}
\sum\limits_{j \in\mathcal{N}:j\neq i}\sum\limits_{k = 1}^{h(j)} k \cdot y^{k}_{ij} - \sum\limits_{j\in\mathcal{T}:j\neq i}\sum\limits_{k = 1}^{h(j)} k \cdot y^{k}_{ji} = 1 \quad  \forall i \in \mathcal{T}
\end{equation}
Constraints (\ref{eqn:c1}) and (\ref{eqn:c2}) enforce \cone.

The set $\chi$ stores pairs of arcs $\left\lbrace (i,j),(u,v)\right\rbrace$, which are crossing each other. Excluding crossing arcs in the solution (\cthree) is ensured by the simultaneous application of the following linear inequalities
\begin{equation} \label{eqn:c3}
x_{ij} + x_{ji} + x_{uv} + x_{vu} \leq 1 \quad  \forall \left\lbrace (i,j),(u,v)\right\rbrace \in \chi
\end{equation}
 \begin{equation} \label{eqn:c4}
\sum\limits_{k = 1}^{h(j)} y^{k}_{ij} - x_{ij} \leq 0 \quad  \forall (i,j) \in \mathcal{A}
\end{equation}
Constraint (\ref{eqn:c5}) represents a set of valid inequalities, initially proposed in \cite{Cerveira2016OptimalCase}, to tighten the mathematical model.
\begin{gather} \label{eqn:c5}
- \sum\limits_{j \in \mathcal{N}:j\neq i}\sum\limits_{k = v + 1}^{h(j)} \floor*{\frac{k - 1}{v}} \cdot y^{k}_{ij} + \sum\limits_{j \in \mathcal{T}:j\neq i}\sum\limits_{k = v}^{h(j)} y^{k}_{ji} \leq 0 \\ \quad \forall v = \left\lbrace 2,\cdots,Q-1\right\rbrace \wedge i \in \mathcal{T} \nonumber
\end{gather}

\section{Computational Experiments}
\label{experiments}
The following experiments have been carried out on an Intel Core i7-6600U CPU running at 2.50 GHz and with 16 GB of RAM. The chosen MILP solver is the branch-and-cut solver implemented in IBM ILOG CPLEX Optimization Studio V12.10 \cite{IBM2021IBMManual}. All routines have been coded in Python 3.7.

\begin{figure}[H]
  \centering \includegraphics[width=0.30\textwidth]{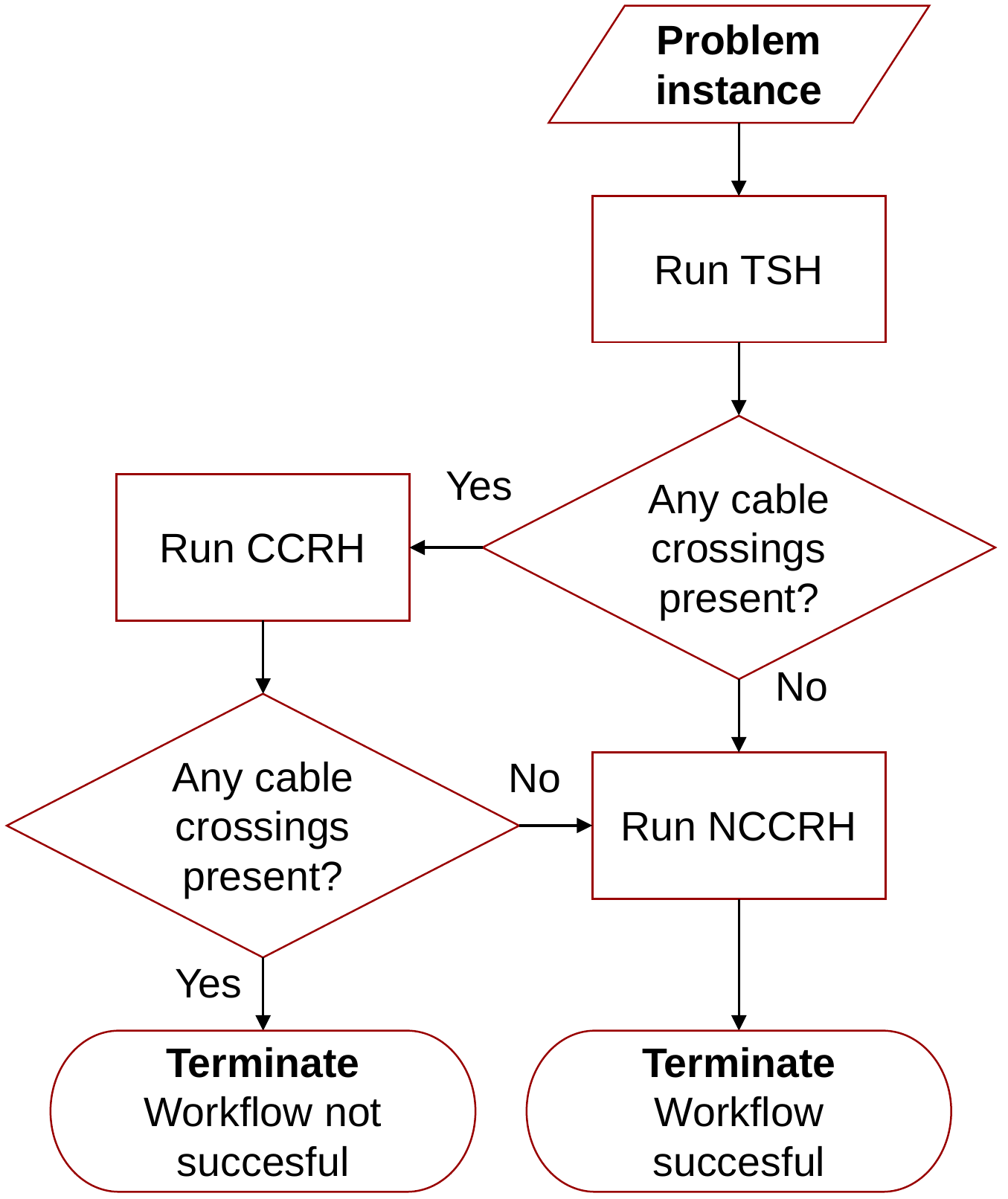}
  \caption{Flowchart of the proposed framework}
  \label{fig:framework}
\end{figure}

The flowchart assembling the algorithms of Section \ref{heuristics} is presented in Fig. \ref{fig:framework}. Given a problem instance, characterized by a fixed WTs and OSSs layout and a set of cables available $\mathcal{C}$ (with capacities $\mathcal{Q}$ and costs $\mathcal{W}$), the task consists in finding a feasible point satisfying \cone, \ctwo, and \cthree.
First, the TSH is run in order to get an initial point, which is then evaluated for the crossing constraint \cthree. Second, In case these are not satisfied, the algorithm CCRH is called, seeking for eliminating those invalidities. Lastly, the retrieved feasible point is intended to be refined by means of the NCCRH.

Apart from the important aspect of getting a feasible point for a NP-Hard problem without external solvers, a not least useful possibility of the proposed framework of  Fig. \ref{fig:framework} (in case of a successful termination), is to warm-start the solver when tackling the global optimization model (Section \ref{global}), utilizing this point. A warm-start is particularly useful for mixed integer problems as the collection system for OWFs, as generally they help the optimizer to activate internal heuristics, causing a faster convergence. The testbed of Table \ref{tab:param_bench} is implemented with the aim of quantifying both functionalities in large-scale real-world cases, extracted from \cite{perez2019global} and \cite{Fischetti2018OptimizingLosses}, plus two extra synthetic OWFs with random pattern of WTs from \cite{perezwes}. For all experiments $\upsilon=15$ (the 15 closest WTs to each WT are included in the input graph $\mathcal{G}$), as this is a reasonable value to cover the global minimum for real-world projects \cite{perez2019global}. Results for the performance of the solver-free heuristics are deployed in Section \ref{res_rep_ref}, and the benefits of warm-starting are presented in Section \ref{warm_starting}. Data related to the real-world OWFs under study are available in \cite{ESCAKIS-ORCA}.
\begin{table}[H]
  \centering
  \captionsetup{justification=centering, labelsep=newline,textfont = sc}
  \caption{Input parameters for computational experiments}
  \label{tab:param_bench}
    \begin{tabular}{ccccc}
    \toprule
    {Instance} & OWF & $n_{\text T}$ & $\mathcal{Q}$ & $\mathcal{W}$ [M\EUR{}/\si{\km}] \\
    \midrule
    1     & \multirow{3}[1]{*}{Horns Rev 1} & \multirow{3}[1]{*}{80} & $\left\lbrace7,11,13\right\rbrace$ & $\left\lbrace0.37,0.39,0.43\right\rbrace$\\
    2     &       &       & $\left\lbrace7,12\right\rbrace$  & $\left\lbrace0.44,0.45\right\rbrace$ \\
    3     &       &       & $\left\lbrace10,14\right\rbrace$ & $\left\lbrace0.44,0.62\right\rbrace$ \\
    4     & \multirow{2}[0]{*}{Ormonde} & \multirow{2}[0]{*}{30} & $\left\lbrace5,10\right\rbrace$  & $\left\lbrace0.41,0.61\right\rbrace$ \\
    5     &       &       & $\left\lbrace4,9\right\rbrace$   & $\left\lbrace0.38,0.63\right\rbrace$ \\
    6     & \multirow{2}[0]{*}{DanTysk} & \multirow{2}[0]{*}{80} & $\left\lbrace4,6,8\right\rbrace$ & $\left\lbrace0.37,0.39,0.43\right\rbrace$ \\
    7     &       &       & $\left\lbrace6,8\right\rbrace$   & $\left\lbrace0.44,0.62\right\rbrace$ \\
    8     & \multirow{2}[0]{*}{Thanet} & \multirow{2}[0]{*}{100} & $\left\lbrace7,15\right\rbrace$  & $\left\lbrace0.38,0.63\right\rbrace$ \\
    9     &       &       & $\left\lbrace7,10\right\rbrace$  & $\left\lbrace0.44,0.62\right\rbrace$ \\
    10    & Random O\newline{} & 74    & $\left\lbrace7,11,13\right\rbrace$ & $\left\lbrace0.37,0.39,0.43\right\rbrace$ \\
    11    & Random I  & 74    & $\left\lbrace4,9\right\rbrace$   & $\left\lbrace0.38,0.63\right\rbrace$ \\
    \bottomrule
    \end{tabular}%
\end{table}%
\subsection{Performance analysis of the solver-free heuristics}
\label{res_rep_ref}
The performances of the TSH, CCRH, and NCCRH algorithms are available in Table \ref{tab:results_table}. Only the problem instances 4, 5, and 8 lead to feasible points when applying the TSH algorithm. For the other instances, up to 15 cable crossings (instance 3, 15cr.) appear after very rapid computing time of this heuristic (in the order of few hundreds of milliseconds). For instances 4, 5, and 8, the feasible points have a solution value greater than the best known solution (after correspondingly solving the global optimization model in each case) by $0.86\%$, $2.04\%$, and $8.87\%$, respectively.

\begin{table*} [!t]
\centering
\captionsetup{justification=centering, labelsep=newline,textfont = sc}
\caption{Performance of the solver-free heuristics}
\label{tab:results_table}
    \begin{tabular}{ccccccccccc}
    \toprule
    \multirow{2}[2]{*}{Inst.} & \multicolumn{3}{c}{TSH} & \multicolumn{3}{c}{CCRH} & \multicolumn{3}{c}{NCCRH} & \multicolumn{1}{p{4.455em}}{Gain with } \\
          & Sol. [M\EUR{}] & Time [\si{\ms}] & Dif. best [\%] & Sol. [M\EUR{}] & Time [\si{\s}] & Dif. best [\%] & Sol. [M\EUR{}] & Time [\si{\s}]-(It) & Dif. best [\%] & \multicolumn{1}{l}{NCCRH [\%]} \\
    \midrule
    1     & Inf-6cr. & 246   & -     & 24.63 & 9.30  & 27.22 & 24.55 & 77.50 (4it) & 26.81 & -0.32 \\
    2     & Inf-10cr. & 262   & -     & 27.66 & 24.22 & 22.50 & 27.65 & 46.40 (1it) & 22.45 & -0.04 \\
    3     & Inf-15cr. & 272   & -     & 32.40 & 33.52 & 37.99 & 32.20 & 41.01  (1it) & 37.14 & -0.62 \\
    4     & 8.18  & 20    & 0.86 & -     & -     & -     & 8.18  & 0.9 (0it) & 0.86 & - \\
    5     & 8.52  & 29    & 2.04 & -     & -     & -     & 8.50  & 1.75 (1it) & 1.80 & -0.23 \\
    6     & Inf-9cr. & 176   & -     & 45.09 & 7.91  & 16.45 & 45.09 & 13.77 (0it) & 16.45 & - \\
    7     & Inf-9cr. & 188   & -     & 58.79 & 6.99  & 19.03 & 58.79 & 14.71 (0it) & 19.03 & - \\
    8     & 24.19 & 186   & 8.87 & -     & -     & -     & 23.24 & 93.76 ( 2it) & 4.59 & -3.93 \\
    9     & Inf-1cr. & 189   & -     & 26.40 & 1.68  & 1.42 & 26.40 & 34.90 (0it) & 1.42 & - \\
    10    & Inf-7cr. & 135   & -     & 54.62 & 2.04  & 15.48 & 52.97 & 53.07 (3it) & 11.99 & -3.02 \\
    11    & Inf-2cr. & 118   & -     & 65.69 & 9.63  & 6.55 & 65.14 & 29.12 (2it) & 5.66 & -0.84 \\
    \bottomrule
    \end{tabular}%
  \label{tab:addlabel}%
\end{table*}
\begin{figure*}[!t]
\centering
  \subfloat[Design after TSH]{\includegraphics[width=0.5\textwidth]{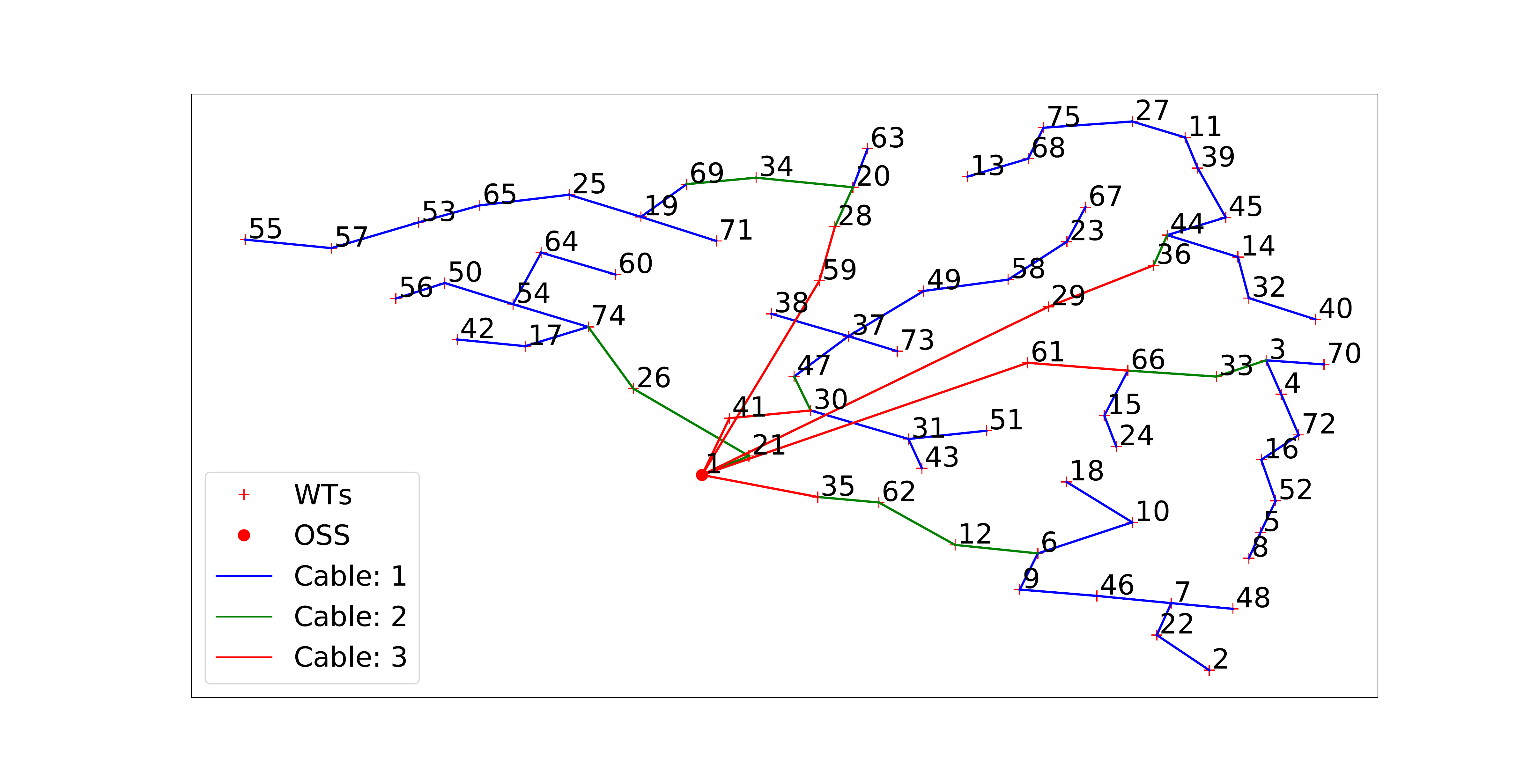}\label{fig:r_o_a}}
  \hfill
  \subfloat[Design after CCRH]{\includegraphics[width=0.5\textwidth]{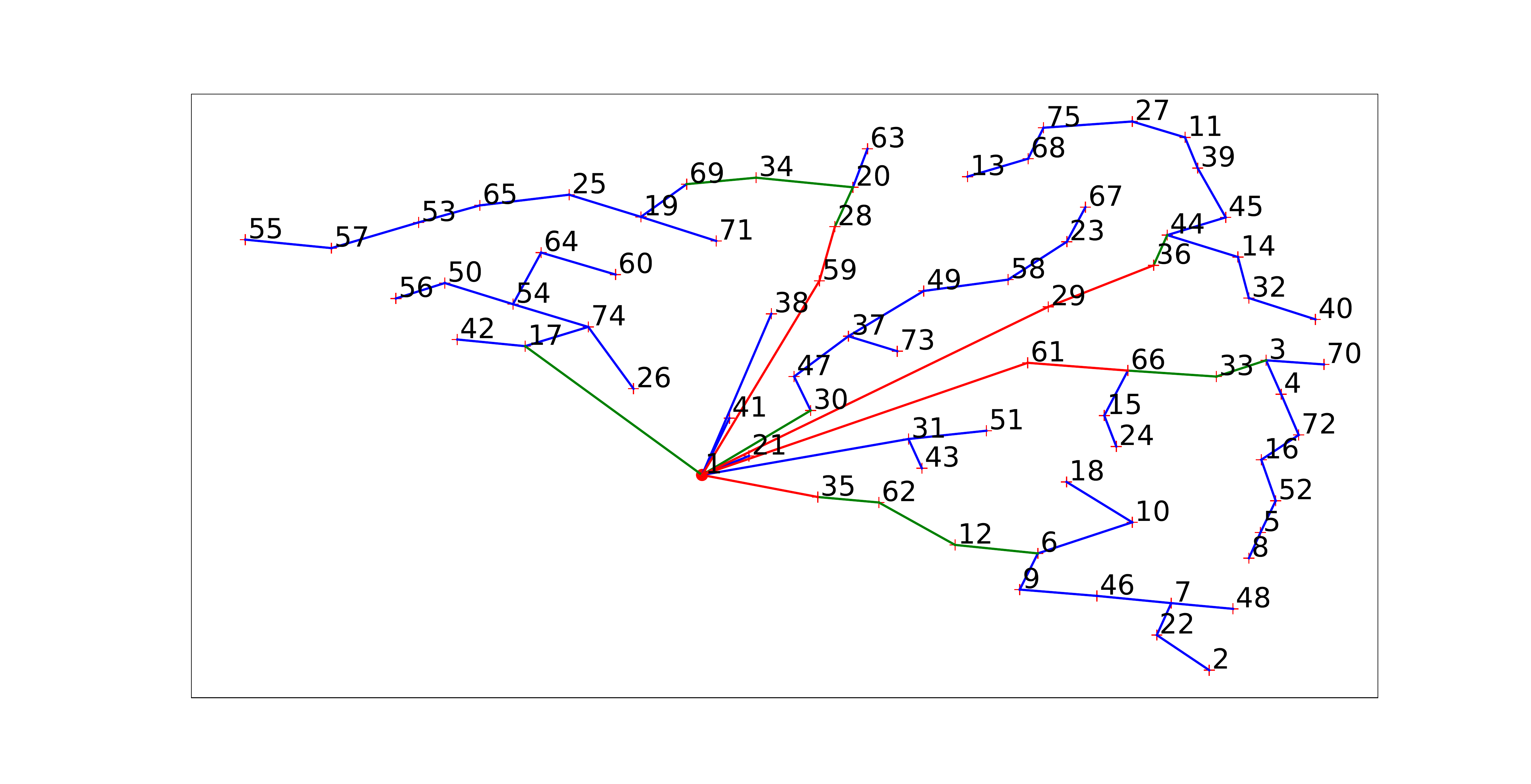}\label{fig:r_o_b}}
  \hfill
\subfloat[Design after NCCRH]{\includegraphics[width=0.5\textwidth]{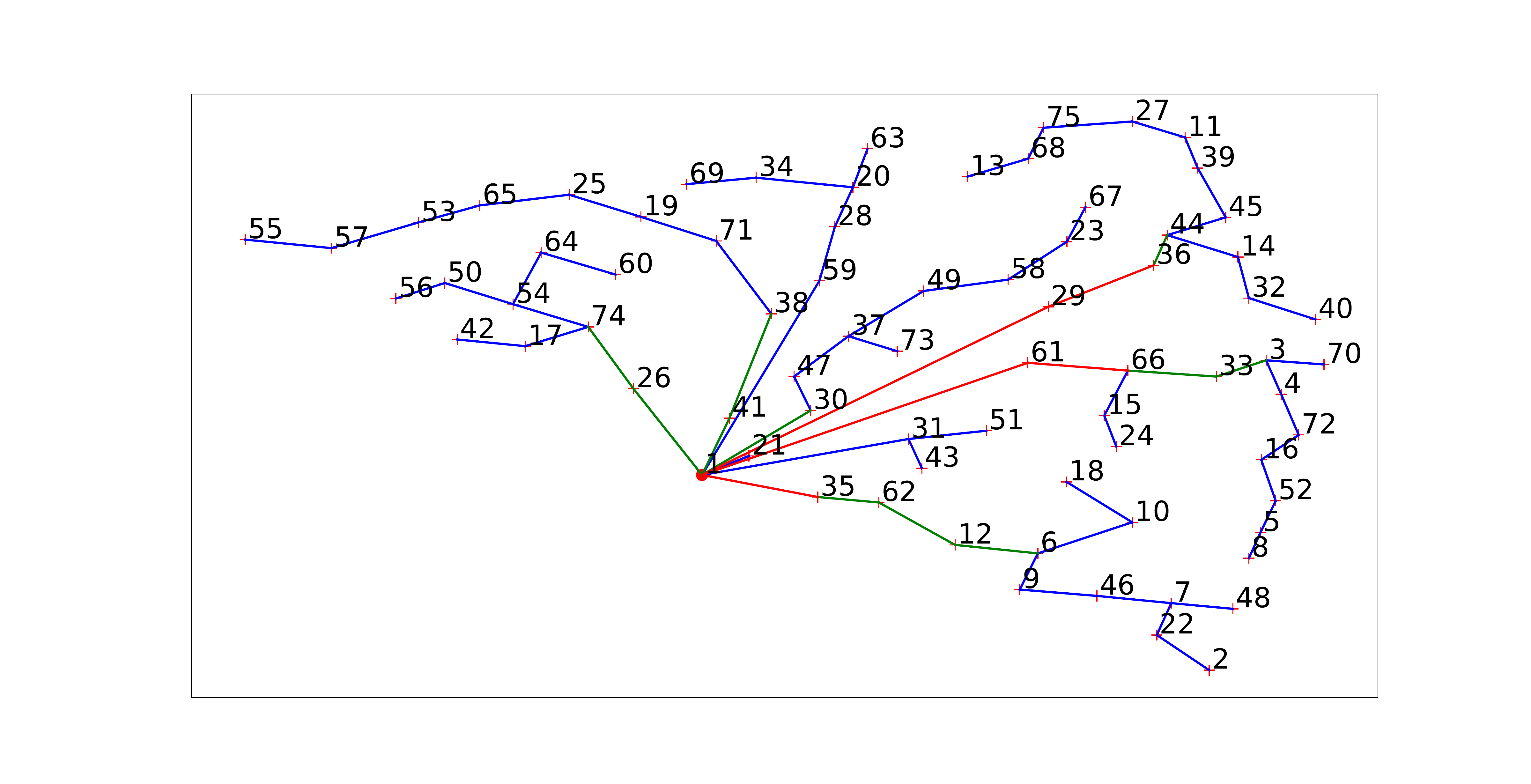}\label{fig:r_o_c}}
  \hfill
  \subfloat[Design after global optimization]{\includegraphics[width=0.5\textwidth]{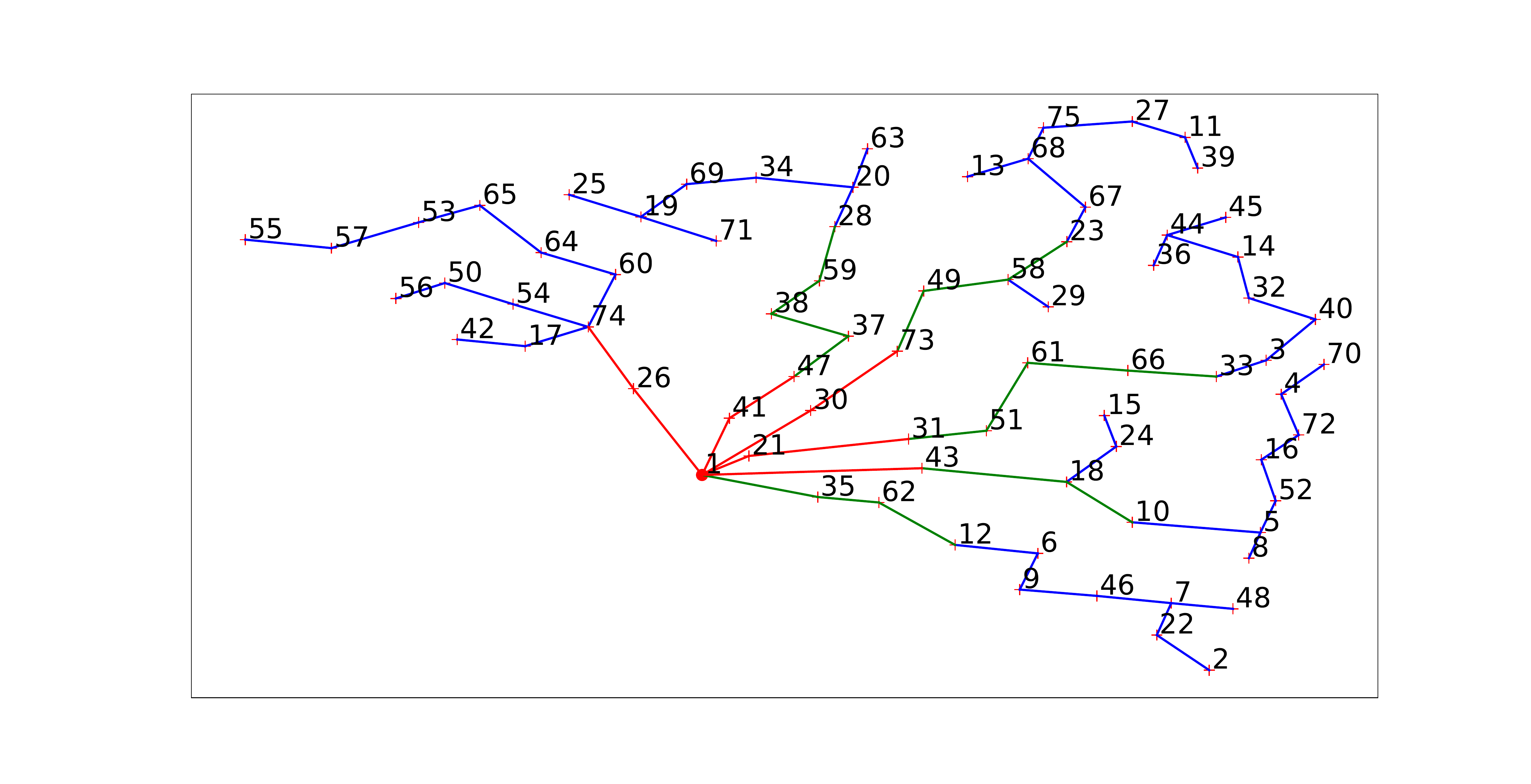}\label{fig:r_o_d}}
\caption{Collection system designs for problem instance 10}
\label{fig:instance10}
\end{figure*}
The CCRH algorithm repairs all the infeasible points after the TSH. The longest computing is of \SI{33.52}{\s} (for problem instance 3, where the greatest number of cable crossings are present) with an overall average of \SI{11.91}{\s}. It is observed a large variation in the percentage deviation with respect to the best known solution; extreme values are $37.99\%$ (also for problem instance 3) and $1.42\%$. It can be appreciated the relation between the number of crossings after the TSH, and the solution quality after the CCRH. The lowest deviations with respect to the best known solution emerge in instances 9 and 11, where the number of crossings are 1 and 2, respectively. This means that the CCRH algorithm is able to fix infeasible points, at expense of moving the solution away of the global minimum.

The NCCRH manages to refine feasible points in $64\%$ of the problem instances. The computing time is in the order of dozens of seconds, with a maximum of \SI{93.76}{\s} (instance 8). The number of successful iterations, i.e., the number of times the flow is improved (line 20 in Algorithm {\ref{alg:alg3}}) is also available (2 iterations for problem instance 8). For the four problem instances where NCCRH does not improve the solution, the computing time spent trying to do so is maximum \SI{35}{\s}. The percentage of improvement after the NCCRH algorithm is presented in the last column of Table \ref{tab:results_table}, with a maximum improvement of $3.93\%$, and average $1.29\%$ for the successful cases. In absolute terms, the improvement is the order of hundreds of thousands of euros in a time scale of seconds.

Graphical results of the designed collection systems for problem instance 10 are illustrated in Fig. \ref{fig:instance10}. The design after the TSH algorithm in Fig. \ref{fig:r_o_a} poses seven cable crossings (observe for example, three crossings with the connection from node 1, OSS, to WT 59), while both constraints \cone\ (note that there is only one electrical path from each WT towards a OSS) and \ctwo\ (each connection meets the capacity set $\mathcal{Q}$) are satisfied. The CCRH algorithm eliminates all crossings while still satisfying \cone\ and \ctwo, see Fig. \ref{fig:r_o_b}. This comes at expense of increasing the accumulated length of cables by $8\%$, implying as well an increase of total investment. The NCCRH succeeds in decreasing investment costs by $3.02\%$, through swapping connections, as for instance, by eliminating connection from 1 to 17 (in Fig. \ref{fig:r_o_b}), and by creating connection from 1 to 26, with subsequent upgrading cable from 26 to 74 (in Fig. \ref{fig:r_o_c}). Ultimately, the best known feasible point is plotted in Fig. \ref{fig:r_o_d}.

\subsection{Benefits of warm-starting for the MILP Model}
\label{warm_starting}
The benefits of warm-starting the MILP solver in terms of solution quality, computing time, and GAP are summarized in Table \ref{tab:ws_benefits}. It is evident that for all aspects, the warm-starting in general helps to get better solutions in shorter times and with tighter certificates.\footnote{The stopping criteria for all experiments using the MILP solver is a GAP less than or equal to $1\%$.} In only one exception (case 6), the warm-starting functionality does not lead to significant improvement.

\begin{figure}[H]
  \centering \includegraphics[width=0.495\textwidth]{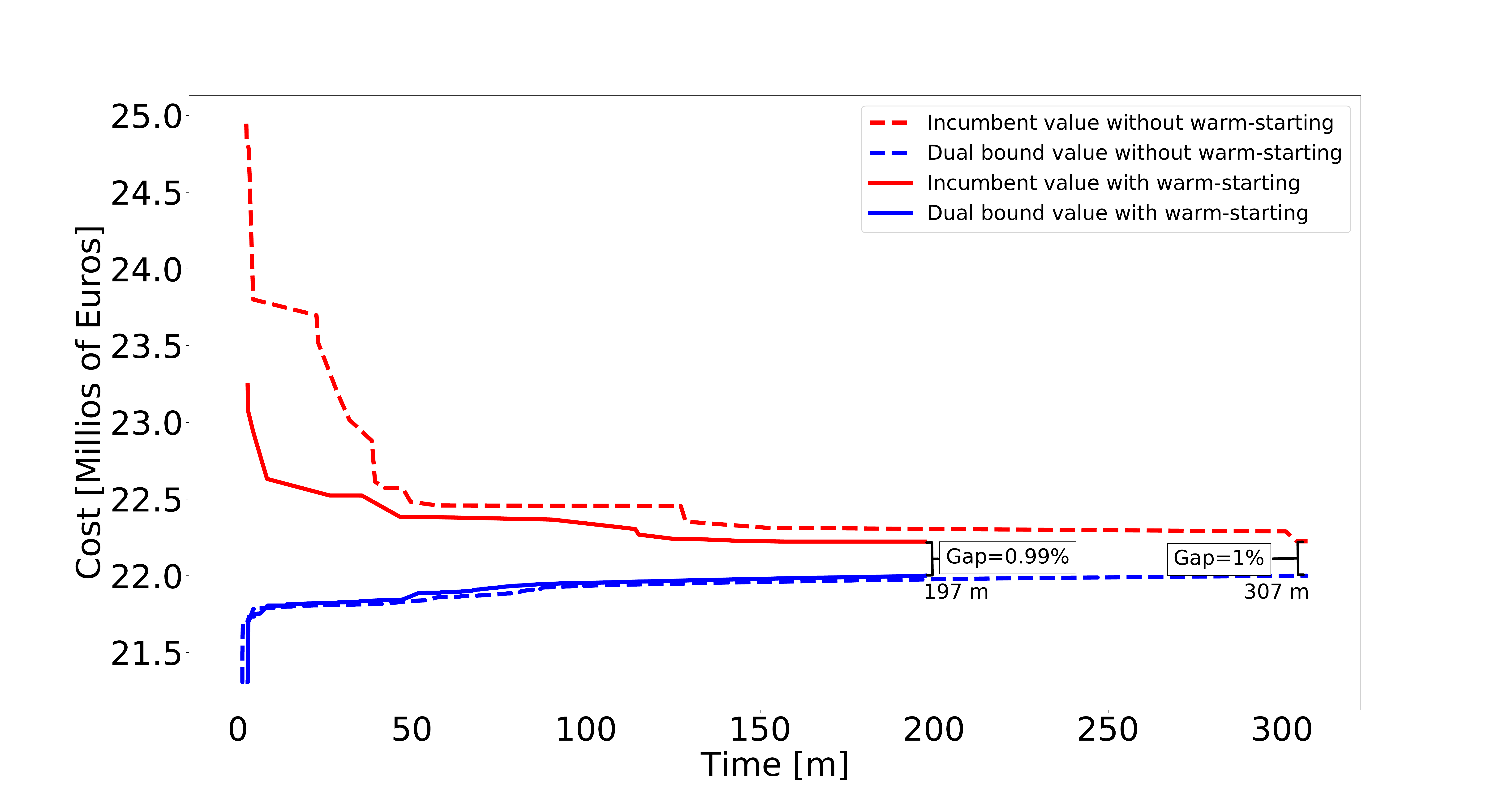}
  \caption{Incumbent and dual bound time evolution without and with warm-starting for problem instance 8}
  \label{fig:ws_nows}
\end{figure}

Problem instance 8 requires \SI{307}{\m} to be solved with the MILP solver without warm-starting. The real-time evolution of the best known feasible point (incumbent) and the best known achievable solution (dual bound) is illustrated in Fig. \ref{fig:ws_nows}. Interestingly, the warm-starting seems to help the solver in such a way that the incumbent curve shifts towards the left, resulting in a faster convergence of \SI{197}{\m}. Oppositely, the dual bound curve appears to be unaffected, which is an indication that supplementary methods should be proposed to speed up its behaviour. 
\begin{table}[H]
  \centering
  \captionsetup{justification=centering, labelsep=newline,textfont = sc}
  \caption{Benefits of warm-starting the MILP solver}
  \label{tab:ws_benefits}
    \begin{tabular}{ccccccc}
    \toprule
    \multirow{2}[1]{*}{I.} & \multicolumn{3}{c}{Global (No warm-starting)} & \multicolumn{3}{c}{Global (With warm-starting)} \\
          & S. [M\EUR{}] & T [\si{\m}] & GAP [\%] & S. I. [\%] & T. I. [\%] & G. I. [\%] \\
    \midrule
    1     & 19.43 & 1.60  & 0.34  & -0.36 & -35.63 & -100.00 \\
    2     & 22.58 & 1.08  & 0.00  & 0.00 & -12.04 & 0.00 \\
    3     & 23.59 & 8.01  & 0.65  & -0.47 & -25.97 & -69.23 \\
    4     & 8.13  & 0.20  & 0.91  & -0.25 & -25.00 & -28.57 \\
    5     & 8.39  & 0.25  & 0.99  & -0.48 & -16.00 & -20.20 \\
    6     & 38.72 & 1.12  & 0.27  & 0.28 & 165.18 & 44.44 \\
    7     & 49.39 & 2.79  & 0.88  & 0.06 & -1.43 & -9.09 \\
    8     & 22.22 & 307   & 1.00  & 0.00 & -35.83 & -1.00 \\
    9     & 26.11 & 2.46  & 0.70  & -0.31 & -56.50 & -10.00 \\
    10    & 47.33 & 15.49 & 0.56  & -0.06 & -10.85 & -25.00 \\
    11    & 61.65 & 10.53 & 0.98  & 0.00  & -37.89 & 1.02 \\
    \bottomrule
    \end{tabular}%
\end{table}%
The improvement ratio for the NCCRH algorithm in instance 8 is equal to $(23.24-24.19)$M\EUR{}/\SI{93.76}{\s}$\approx-0.01$M\EUR{}/\si{\s}. The values of this coefficient after \SI{4}{m} for both incumbent curves in Fig. \ref{fig:ws_nows},\footnote{Computing time for incumbent curves stabilization with solutions within $7\%$ to the best know feasible point.} are actually considerably lower than this value. This proves the strong contribution of this heuristic, which is indeed able to refine solutions in competitive computing times.
\section{Conclusion}
\label{conclusion}
The proposed method provides a workflow of solver-free heuristics to address the OWF collection system problem. 

This manuscript brings along two main contributions. First, the ability to empirically get feasible points for large-scale real-world OWF collection system, within computing time in the order of dozen of seconds, and solution quality with a deviation between $0.86\%$ to $37.14\%$ to the best known feasible point. On the first hand, the CCRH algorithm behaves satisfactorily for all the studied problem instances. On the other hand, the NCCRH improves the solution in $64\%$ of the cases, with an enhancement up to $3.93\%$. Second, the retrieved feasible points are fed into the branch-and-cut solver as a warm-starting solution, which demonstrate the benefit to accelerate convergence (up to $56.50\%$ of time reduction), and to actually come up with slightly better final solutions.

The NCCRH is inspired in previous works where the classic minimum cost flow algorithm is modified, taking into consideration the particular properties of the OWF collection system problem. The finite capacity of cables, along with their step cost function, and typical engineering constraints verification, are incorporated when finding negative cycles that lead to cost minimization.

Future work can focus on: (i) different initial feasible flows to provide to the NCCRH algorithm, (ii) strategies to account for more then one cable swapping simultaneously, (iii) modification of Bellman-Ford algorithm to track longer negative cycles, while satisfying the constraints, and (iv) strategies to not only improve convergence of the incumbent curve, but also the dual bound curve.

\section*{Acknowledgment}
This research has received funding from the TOPFARM CCA project (\url{https://topfarm.pages.windenergy.dtu.dk/TopFarm2/}).
The author thanks Prof. Mathias Stolpe and Prof. Nicolaos A. Cutululis for their support, inputs, and feedback on the manuscript.

\section*{Conflicts of interest}
No conflicts of interest are present.

\ifCLASSOPTIONcaptionsoff
  \newpage
\fi
\bibliographystyle{IEEEtran}
\bibliography{main.bbl}
\vskip -2\baselineskip plus -1fil
\begin{IEEEbiography}
[{\includegraphics[width=1in,height=1.25in,clip,keepaspectratio]{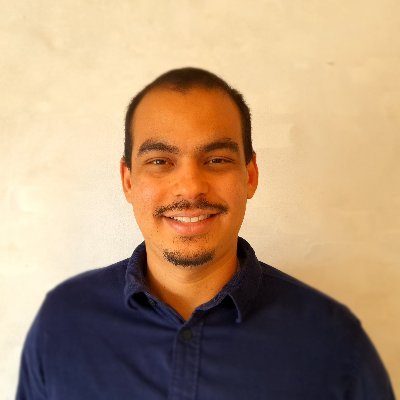}}]{Juan-Andr\'es P\'erez-R\'ua}  received the B.Sc. degree (SCL Hons) in Electrical Engineering from the Technological Univ. of Bolivar, Colombia (2012), the M.Sc. degree in Electrical Power Systems from the ISEC college, Portugal, the Univ. of Nottingham, England, and the Univ. of Oviedo, Spain (2016), and the Ph.D. degree from DTU Wind Energy, Denmark (2020). His current research interest lies in multidisciplinary optimization of renewable energy systems, with special focus on electrical network design and operation of wind and solar plants.
\end{IEEEbiography}
\end{document}